\documentclass[aps,prd,preprint,groupedaddress,superscriptaddress,nofootinbib]{revtex4}%
\usepackage[utf8]{inputenc}
\usepackage[T1]{fontenc}
\usepackage{soul} 
\usepackage{amsmath}
\usepackage{graphicx}
\usepackage{xcolor}

\begin{document}%

\title{\textbf{Perturbative treatment of the luminosity distance}}

\author{Dimitar Ivanov}
\email{divanov@sissa.it}

\affiliation{SISSA - International School for Advanced Studies, 
Via Bonomea 265, 34136 Trieste TS, Italy.}

\affiliation{INFN, Sezione di Trieste, Via Valerio 2, 34127 Trieste TS, Italy.}

\author{Stefano Liberati}
\email{liberati@sissa.it}

\affiliation{SISSA - International School for Advanced Studies, 
Via Bonomea 265, 34136 Trieste TS, Italy.}

\affiliation{INFN, Sezione di Trieste, Via Valerio 2, 34127 Trieste TS, Italy.}

\author{Matteo Viel}
\email{viel@sissa.it}

\affiliation{SISSA - International School for Advanced Studies, 
Via Bonomea 265, 34136 Trieste TS, Italy.}

\affiliation{INFN, Sezione di Trieste, Via Valerio 2, 34127 Trieste TS, Italy.}

\affiliation{INAF, Osservatorio Astronomico di Trieste, 
                    via Tiepolo 11, I-34131 Trieste, Italy.}

\author{Matt Visser}

\email{matt.visser@sms.vuw.ac.nz}

\affiliation{School of Mathematics and Statistics, Victoria University of Wellington; \\
PO Box 600, Wellington 6140, New Zealand}

\bigskip
\begin{abstract}

We derive a generalized luminosity distance \emph{versus} redshift relation for a linearly perturbed FLRW (Friedmann--Lemaitre--Robertson--Walker) metric with two scalar mode excitations. We use two equivalent approaches, based on the Jacobi map and the van Vleck determinant respectively. We apply the resultant formula to two simple models --- an exact FLRW universe and an approximate FLRW universe perturbed by a single scalar mode sinusoidally varying with time. For both models we derive a cosmographic expansion for $d_L$ in terms of $z$. We comment on the interpretation of our results and their possible application to more realistic cosmological models.

\end{abstract}



\maketitle

\tableofcontents



%
%


\section{Introduction}

Supernovae observations suggest that the universe is currently undergoing a period of accelerated expansion~\cite{Planck}. A crucial assumption in the interpretation of these results is that our universe is homogeneous and isotropic on sufficiently large scales, \emph{i.e.}, that the background is at least approximately FLRW so that
\begin{equation}
\mathrm{d}s^2 \approx - \mathrm{d}t^2 + a^2(t)
\left[\frac{\mathrm{d}r^2}{1-kr^2} + r^2 \mathrm{d} \Omega^2\right].
\end{equation}

For the purposes of this article we shall immediately set $k=0$ as we feel that there are both good theoretical motivations and observational evidence for that choice~\cite{Planck2}. 
(Though see~\cite{Ratra} for a recent countervailing point of view.)
In a recent related article on non-perturbative aspects of the luminosity distance~\cite{Non-perturbative} we were careful to retain potentially nonzero values of $k$. In the current article we are ultimately interested in perturbative analyses, and it makes sense to set
\begin{equation}
\mathrm{d}s^2 \approx - \mathrm{d}t^2 + a^2(t)
\left[{\mathrm{d}r^2} + r^2 \mathrm{d} \Omega^2\right].
\end{equation}
Given this assumption, the most straightforward way of analyzing the supernova data is via a cosmographic approach~\cite{Visser, Visser:jerk, Cattoen:2007, Cattoen:2007a, Cattoen:2007b, Cattoen:2008, Visser:2009, Vitagliano} --- in FLRW cosmology one can, independently from the gravitational field equations, express the luminosity distance of a standardizable candle as a power series of its redshift~\cite{Visser,Visser:jerk}.In the absence of any peculiar velocities, and expanding around the current epoch, for an exact FLRW universe one has
\begin{align}
d_L(z) &= \frac{z}{H_0} \Big\{ 1 + \frac{1}{2} \Big[1-q_0\Big]z - \frac{1}{6}\Big[1-q_0 - 3q_0^2 + j_0\Big]z^2  
\nonumber \\
&+ \frac{1}{24}\Big[2-2q_0 - 15q_0^2 - 15q_0^3 + 5j_0 + 10q_0j_0 +s_0\Big]z^3 + O(z^4) \Big\}.
\label{jackson}
\end{align}
Here the cosmographic coefficients --- Hubble rate, deceleration parameter, jerk, and snap, are defined respectively in terms of $t$-time derivatives as
\begin{equation}
H = \frac{\dot{a}}{a}; \qquad
q = - \frac{1}{H^2} \frac{\ddot{a}}{a};
\qquad
j = \frac{1}{H^3} \frac{\dddot{a}}{a};
\qquad 
s = \frac{1}{H^4} \frac{\ddddot{a}}{a}.
\label{e:H_t}
\end{equation}
Given enough supernovae observations one can constrain the shape of the cosmographic curve $d_L(z)$ and thus constrain the values of the cosmographic parameters. Current constraints suggest that $q_0 < 0$~\cite{Planck,Planck2}, which justifies the claim that the universe is currently in a phase of an accelerated expansion. In general, for a perturbed FLRW universe a cosmographic analysis along these lines, or along the lines indicated below, will only provide \emph{part} of the full formula for the luminosity distance, and in this article we shall among other things analyze various deviations from simple cosmography.

Traditionally, the accelerated expansion is explained by assuming an unknown matter component with negative pressure which enters the right-hand side of Einstein's field equation. This matter component is usually assumed to take the form of a cosmological constant or vacuum energy and thus to be constant over space and time. However, there exist a plethora of models where this so called dark energy varies with time, and might potentially also vary with space~\cite{EFTDE, Everpresent, Caldwell, Mukhanov, Li}. There also exist alternative explanations for the observation $q_0 < 0$, such as modification of GR at cosmological scales~\cite{Sotiriou:2008,Lobo:2008,DeFelice:2010}, and significant departures from exact FLRW cosmology~\cite{Buchert, Ellis, RelCosm,CFLRW,Visser:2015-Buchert}. In the case of significant departures from homogeneity or/and from uniform dark energy, one does not expect the theoretical relation \eqref{jackson} to hold any more, and one has to perform the supernovae data fitting with some sort of improved $d_L(z)$ relation.

In this paper we derive a generalised $d_L(z)$ relation and consider its implications. Our motivation for this is twofold. On the one hand we want to allow for the possibility to fit supernovae data with alternative cosmological models with varying dark energy, and thus constrain the parameter space of such models. On the other hand, we want to consider the implications of inhomogeneities due to the large scale structure of the universe on the interpretation of the supernovae results.

There have been numerous attempts to derive a generalised $d_L(z)$ relation ever since the paper of Sasaki~\cite{Sasaki}. In that paper, under suitable conditions, the following formula for the luminosity distance in a perturbed geometry was derived:
\begin{equation}
d_L(z, \lambda_s) = \bar{d}_L(z) \left[ 1 + \Big(\frac{a^{\prime}}{a}\delta \eta \Big)_o 
+ \coth \left(\sqrt{-k} \lambda_s\right) \sqrt{-k} \delta \lambda_s - \frac{1}{2} \int^{\lambda_s}_0 \delta \theta (\lambda) d \lambda \right].
\end{equation}
Here $\bar{d}_L(z)$ is the luminosity distance evaluated at the background, while $\delta \eta$, $\delta \lambda$ and $\delta \theta$ are the perturbations of the conformal time, the affine parameter and the expansion. Further progress was made in~\cite{Durrer}. Their expression $(53)$ bears close similarity to our expression~\eqref{lumdis}. Generalised formulas for $d_L$ (or some function of it, such as the magnitude or the fractional fluctuation) have also been derived in~\cite{Riotto, Umeh:2012, Yoo, Gasperini, BenDayan}. In~\cite{Yoo} the authors compute the two-point correlation function of the luminosity distance while in~\cite{BenDayan} the authors compute the luminosity distance to second order in perturbations in the geodesic lightcone gauge and then transform to the Poisson gauge.

In this paper we shall assume the universe is well described by a linearly perturbed FLRW metric with two scalar mode excitations
\begin{equation}
\mathrm{d} s^2 = a^2(\eta) 
\Big[-(1+2\Psi)\mathrm{d}\eta^2 + \delta_{ij} (1+2\Phi) \mathrm{d}x^i \mathrm{d}x^j\Big] .
\label{robert}
\end{equation}
Here the conformal time coordinate $\eta$ is defined as $\mathrm{d}\eta = \frac{\mathrm{d}t}{a(t)}$. We derive a formula for the luminosity distance in this geometry using two different but closely related approaches --- the Jacobi map approach and the van Vleck determinant approach. Both approaches are kinematic in nature --- they assume nothing about what the correct theory of gravity is. While the Jacobi map calculation is similar to the one performed in~\cite{Durrer}, the van Vleck determinant calculation is entirely new and, as we will see, leads to the same final formula for the luminosity distance. We rewrite this final formula in terms of the various contributions to the redshift to the extent possible. We emphasise the cosmographic approach by first reviewing the cosmographic expansion in FLRW universe and then by performing a generalised cosmographic expansion for a simple toy model with a sinusoidally varying scalar perturbation. We also show how to systematically introduce Doppler redshifts in the cosmographic series.

The structure of this paper is the following. In section \ref{S:cosmo}  we discuss cosmographic generalities and in section \ref{S:luminosity} we introduce the formalism behind the two approaches and verify that they reproduce the correct result in a FLRW universe. Furthermore, we show how to adapt the formalism to get a handle on peculiar Doppler shifts in a FLRW universe. In section \ref{S:linear} we introduce linear perturbations to the FLRW metric and derive formulas for the redshift and luminosity distance in terms of conformal time  using the two approaches. In section \ref{S:toy} we apply the derived formulas to a simple toy model and show how a generalized cosmographic expansion can be obtained in this case. We discuss the implications of our results  and conclude in section \ref{S:summary}.

Throughout the paper we use units in which $c=1$ and the spacetime metric is taken to have a signature $(-1, 1, 1, 1)$. 

\section{Cosmographic generalities}
\label{S:cosmo}

Cosmographic analyses make good physical sense whenever the cosmological spacetime can be sliced by spacelike hypersurfaces which can be factored into an overall ``size of the universe''  (depending only on some convenient global time parameter $t$, possibly some proper time measured by some class of fiducial observers) multiplied by something that depends on the ``shape'' of the spatial slices. That is, take
\begin{equation}
ds^2 =  - N(t,\vec x)^2 \, dt^2 + a(t)^2 \, [g_{shape}(t,\vec x)]_{ij} \, dx^i dx^j.
\end{equation}
This form of the metric is a variant on the notion of   ``synchronous gauge''.  It might be called ``pre-synchronous'', or ``conformally synchronous'', and is sufficiently general to be compatible with our two-mode ansatz as presented in equation~\eqref{robert}.\footnote{Observe that the phrase ``synchronous gauge'', where $N(t,\vec x)=1$, is somewhat of a misnomer. When enforced globally it enforces the existence of a timelike geodesic vorticity-free congruence $V=dt$. The ``conformally synchronous'' gauge is less restrictive, only requiring the existence of a timelike vorticity-free congruence $V= N^{-1} \, dt$, that is not necessarily geodesic.
Note we also want $\partial_t \det([g_{shape}(t,\vec x)]_{ij})$ to be perturbatively small. }

Whenever such a decomposition makes sense one can further construct a ``conformal time'' coordinate $d\eta = dt/a(t)$ and use this to recast the spacetime metric as
\begin{equation}
ds^2 =  a(\eta)^2 \left\{ - N(\eta,\vec x)^2 \, d\eta^2 +\, [g_{shape}(\eta,\vec x)]_{ij} \, dx^i dx^j \right\}.
\end{equation}
As long as this can be done (and this is a rather mild constraint on the cosmology), one can undertake a cosmographic analysis either in terms of the $t$-time derivatives, [as in equation (\ref{e:H_t}) above], or in terms of $\eta$-time derivatives
\begin{equation}
\label{e:H_eta}
\mathcal{H} = \frac{a^{\prime}}{a}; \qquad
\mathcal{Q} = -\frac{1}{\mathcal{H}^2} \frac{a^{\prime \prime}}{a}; \qquad  
\mathcal{J} = \frac{1}{\mathcal{H}^3} \frac{a^{\prime \prime \prime}}{a}.
\end{equation}
Indeed, we can expand the scale factor in a truncated Taylor series around the ``observer'' conformal time $\eta_o$, the conformal time equivalent of the present epoch, so that
\begin{align}
a(\eta) &= a(\eta_o) \bigg[1 + \mathcal{H}_o (\eta - \eta_o) - \frac{\mathcal{H}_o^2 \mathcal{Q}_o}{2} (\eta - \eta_o)^2 
+ \frac{\mathcal{H}_o^3 \mathcal{J}_o}{6} (\eta - \eta_o)^3 + O(\eta - \eta_o)^4 \bigg],
\end{align}
and, using $1+z=a_o/a(\eta)$, we can derive an expansion of $z$ in terms of $\mathcal{H}_o (\eta - \eta_o)$. We find
\begin{align}
z(\eta) &= -[\mathcal{H}_o (\eta - \eta_o)] + \frac{2+\mathcal{Q}_o}{2} [\mathcal{H}_o (\eta - \eta_o)]^2 
- \frac{\mathcal{J}_o + 6\mathcal{Q}_o + 6}{6} [\mathcal{H}_o (\eta - \eta_o)]^3 
\nonumber\\ &
\qquad + O\Big( [\mathcal{H}_o (\eta - \eta_o)]^4\Big).
\end{align}
Reverting the series, we obtain
\begin{align}
\label{E:cosmo}
\mathcal{H}_o (\eta - \eta_o) = &-z + \frac{2+\mathcal{Q}_o}{2} z^2 
 -\frac{3\mathcal{Q}_o^2 + 6\mathcal{Q}_o + 6 - \mathcal{J}_o}{6} z^3 + O(z^4).
\end{align}
\emph{Mutatis mutandis} there is a completely analogous result in terms of the $t$-time
\begin{align}
{H}_o (t - t_o) = &-z + \frac{2+q_o}{2} z^2 
 -\frac{3q_o^2 + 6q_o + 6 - j_o}{6} z^3 + O(z^4).
\end{align}

Such perturbative expansions can in principle be carried out to arbitrarily high order, and their usefulness is limited only by the extent to which we can measure, estimate, or theoretically predict the Hubble, deceleration, jerk, and higher-order parameters. 
Perhaps the key point is that these cosmographic series make sense under very generic conditions, whenever one is able to peel off an ``overall size'' and a natural ``global time'' for the universe. 
These cosmographic series will generically only be part of the full analysis, (for instance they ignore peculiar velocities and the effect of  local clumping), 
but if the ``overall size'' $a(t)$ or equivalently $a(\eta)$ is chosen appropriately, they can easily be the dominant feature contributing to  the luminosity distance.

\section{The luminosity distance}\label{S:luminosity}
\subsection{Definition and interpretation}
We now consider a spacetime $(\mathcal{M}, g_{\mu \nu})$ and a point source emitting light at the source event $S$. 
An extended observer located at $O$ receives the light emitted by $S$. The intrinsic luminosity of $S$ is related to the flux $F$ measured by $O$ by the integral~\cite{RelCosm}
\begin{equation}
L = \int_{S^2} (1+z)^2 F \; \mathrm{d} A.
\label{int}
\end{equation}
Here $S^2$ is the 2-sphere centred at the source $S$, and passing through the observer $O$, while $z$ is the redshift of the light. If the source radiates isotropically, we can write \eqref{int} as a differential relation
\begin{equation}
F \; \mathrm{d}A_o = \frac{L}{4\pi} \frac{\mathrm{d}\Omega_s}{(1+z)^2}.
\label{diff}
\end{equation}
Here $\mathrm{d}A_o$ is an area element at the observer and $\mathrm{d}\Omega_s$ is the infinitesimal solid angle at the source. The luminosity distance between the source and the observer is defined as
\begin{equation}
d_L(S, O) := \sqrt{\frac{L}{4\pi F}}.
\end{equation}
One can easily see that in a Minkowski spacetime this reduces to the standard notion of distance. 
Using \eqref{diff} the luminosity distance can be written as
\begin{equation}
d_L = (1+z)\, \sqrt{\frac{\mathrm{d}A_o}{\mathrm{d}\Omega_s}}.
\end{equation}
We want to relate the quantity $\frac{\mathrm{d}A_o}{\mathrm{d}\Omega_s}$ to the metric and thus compute the luminosity distance.

\begin{figure}[h!]
\centering
\includegraphics[width=0.5\textwidth]{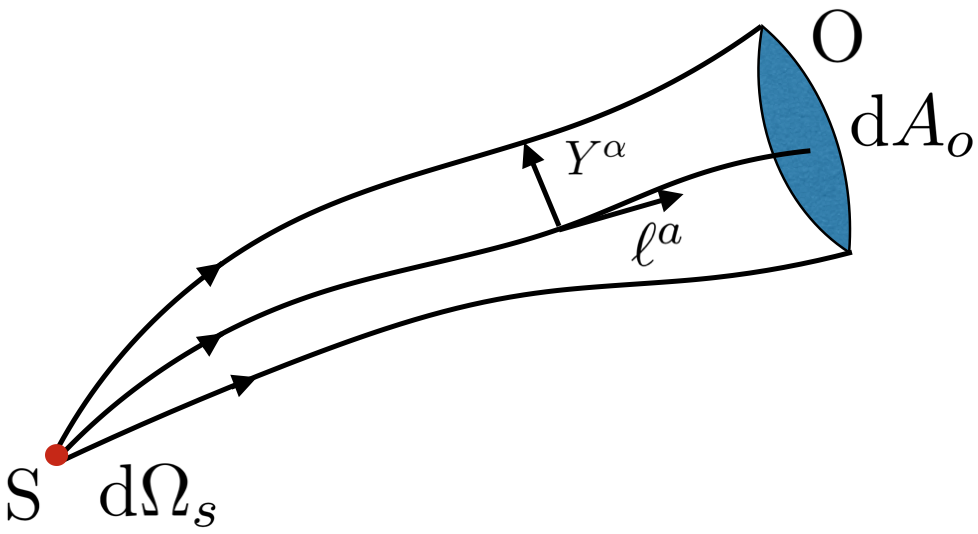}
\caption{\label{F:one}A congruence of light rays emitted at a point source S and received by an extended observer O. The luminosity distance between S and O is given by $d_L = (1+z)\sqrt{\mathrm{d}A_o/\mathrm{d}\Omega_s}$, while $\ell^{\alpha}$ is a tangent vector to a geodesic in the congruence while $Y^{\alpha}$ is a transverse vector connecting different geodesics of the congruence. }
\end{figure} 

\subsection{The Jacobi map and the Jacobi determinant}
\label{Jmd}

The light rays emitted by the source form a congruence of null geodesics (see figure~\ref{F:one}) that can be parametrised as
\begin{equation}
x^{\alpha} = f^{\alpha}(\lambda, y_i).
\end{equation}
Here $\lambda$ is the affine parameter along each light ray, and the $y_i$ parametrise neighbouring rays. For our purposes it is enough to concentrate on a single one-parameter family of light rays
\begin{equation}
x^{\alpha} = f^{\alpha}(\lambda, y).
\end{equation}
The tangent vector and the wave vector are defined as
\begin{equation}
\ell^{\alpha} = \frac{\partial f^{\alpha}}{\partial \lambda};
\qquad
k^{\alpha} = \tilde{\omega} \, \ell^{\alpha}.
\end{equation}
Here $\tilde{\omega}$ is just a constant with dimension $[L^{-1}]$. The geodesic deviation vector is defined as
\begin{equation}
Y^{\alpha} = \frac{\partial f^{\alpha}}{\partial y}.
\end{equation}
For a point source all light rays intersect at S and therefore $Y^{\alpha}_s=0$. The geodesic deviation equation for the family of geodesics is
\begin{equation}
\frac{D^2 Y^{\alpha}}{d \lambda^2} = R^{\alpha}{}_{\beta \gamma \delta} \; \ell^{\beta} \ell^{\gamma} \;Y^{\delta}. \label{geoddev}
\end{equation}
The equation is linear and therefore the solution at $O$ is a linear combination of the initial values at the source $S$ (this is a nontrivial result --- see for instance~\cite{Abramo}). Since $Y^{\alpha}_s=0$, we must have
\begin{equation}
Y^{\rho}_o = \mathcal{J}^{\rho}_{\alpha} (O, S) \frac{DY^{\alpha}_s}{d \lambda},
 \label{sol}
\end{equation}
where $\mathcal{J}^{\rho}_{\alpha} (O, S)$ is called the Jacobi map. It is useful to define the following infinitesimal vectors
\begin{equation}
\delta x^{\alpha}:= Y^{\alpha} \,\delta y \label{inft1},
\end{equation}
which can be thought of as pointing from one geodesic to a neighbouring one along the family,  and
\begin{equation}
\delta \theta^{\alpha} :=  \frac{D Y^{\alpha}}{d \lambda} \, \delta y,
 \label{inft2}
\end{equation}
which connects one geodesic to a neighbouring one at the source and whose magnitude is the angular separation between the two geodesics at the source. Then \eqref{sol} becomes
\begin{equation}
\delta x^{\mu}_{o} = \mathcal{J}^{\mu}_{\alpha} (O, S) \; \delta \theta^{\alpha}_{s}.
 \label{jac1}
\end{equation}
Thus the Jacobi map maps initial directions around the source to vectors transversal to the photon beam at the observer position.

The Jacobi map defined in \eqref{jac1} is a $4$-dimensional map from the tangent space $T_s(\mathcal{M})$ to $T_o(\mathcal{M})$. However the vectors $\delta x^{\mu}_{o}$ and $\delta \theta^{\alpha}_{s}$ live in $2$-dimensional subspaces of the tangent spaces at $O$ and $S$, normal to the four-velocities of the observer and the source, $U_o$ and $U_s$, respectively, and normal to the photon direction at $O$ and $S$ (see Appendix \ref{A:1}). To find the true Jacobi map, we need to project onto these subspaces
\begin{equation}
J(O,S) := P_o \mathcal{J} P_s,
 \label{jac2}
\end{equation}
where $P_o$ and $P_s$ are the projectors
\begin{equation}
(P_s)^{\mu}_{\nu} = (\delta^{\mu}_{\nu} + U^{\mu} U_{\nu} - n^{\mu} n_{\nu})_s 
\qquad
(P_o)^{\mu}_{\nu} = (\delta^{\mu}_{\nu} + U^{\mu} U_{\nu} - n^{\mu} n_{\nu})_o \label{proj2};
\end{equation}
and where $n_s$ and $n_o$ are normalised spacelike vectors pointing in the photon direction in the reference frames of the source and the observer
\begin{equation}
n_o = \big(\ell + (\ell\cdot U)U\big)_o; 
\qquad
n_s = \big(\ell + (\ell\cdot U)U\big)_s. \label{phodir2}
\end{equation}
Here $J(O,S)$ is a 2-dimensional map from a subspace of $T_s(\mathcal{M})$ to a subspace of $T_o(\mathcal{M})$. It follows from its definition in \eqref{jac1} and \eqref{jac2} that 
\begin{equation}
\frac{\mathrm{d}A_o}{\mathrm{d}\Omega_s} = |\det J(O,S)\,|.
\end{equation}
See general discussion in~\cite{Non-perturbative}, and other sources~\cite{Durrer, Riotto, Yoo, Gasperini, Umeh:2012, Umeh:2014, DiDio:2016, BenDayan}.
Thus the luminosity distance is given by
\begin{equation}
d_L(S,O) = |\det J(O,S)|^{\frac{1}{2}} (1+z). \label{lumjac}
\end{equation}
We now want to solve the geodesic deviation equation \eqref{geoddev} in order to find the Jacobi map. In practice it is easier to transform the geodesic deviation equation, which is a 2nd-order differential equation for $Y^{\mu}$, into two coupled 1st-order differential equations for $\delta x^{\mu}$ and $\delta \theta^{\mu}$. (This is similar to what one does in Hamiltonian mechanics where a single 2nd-order differential equation for a given dynamical variable is transformed into two 1st-order differential equations for the dynamical variable and its conjugate momentum.) It follows from \eqref{inft1}, \eqref{inft2} and \eqref{geoddev} that
\begin{equation}
\frac{\mathrm{D} (\delta x^{\mu})}{\mathrm{d} \lambda} = \delta \theta^{\mu};
\qquad\qquad
\frac{\mathrm{D} (\delta \theta^{\mu})}{\mathrm{d} \lambda} = R^{\mu}_{\nu \alpha \beta} \,\ell^{\nu} \ell^{\alpha} \delta x^{\beta}.
\end{equation}
Equivalently
\begin{equation}
\frac{\mathrm{d}(\delta x^{\alpha})}{\mathrm{d} \lambda} = C^{\alpha}_{\nu} (\lambda) \,\delta x^{\nu} + \delta \theta^{\alpha}; 
\qquad\qquad
\frac{\mathrm{d}(\delta \theta^{\alpha})}{\mathrm{d} \lambda} = A^{\alpha}_{\nu} (\lambda) \,\delta x^{\nu} + C^{\alpha}_{\nu} (\lambda) \, \delta \theta^{\alpha}; \label{de2}
\end{equation}
where
\begin{equation}
C^{\alpha}_{\nu} (\lambda) := - \Gamma^{\alpha}_{\mu \nu} \ell^{\mu};
\qquad\qquad
A^{\alpha}_{\nu} (\lambda) := R^{\alpha}_{\rho \mu \nu} \ell^{\rho} \ell^{\mu}.
\end{equation}
In order to find the Jacobi map, this system must be solved consistently with the initial conditions
\begin{equation}
\delta x^{\alpha} (\lambda_s) = 0; 
\qquad\qquad
(\ell^{\alpha} \delta \theta_{\alpha}) (\lambda_s) = (U^{\alpha}_s \delta \theta_{\alpha}) (\lambda_s)=0. \label{ic2}
\end{equation}
Similar equations have also been derived in~\cite{Durrer}.

\subsection{The van Vleck determinant}
The van Vleck determinant measures the deviation from the inverse square law~\cite{vanVleck}
\begin{equation}
F = \frac{L \Delta_{vV}}{4\pi (\lambda_o - \lambda_s)^2 (1+z)^2}.
\end{equation}
This implies that the luminosity distance is given by
\begin{equation}
d_L = (\lambda_o - \lambda_s) (1+z) \Delta_{vV}^{-\frac{1}{2}}.
 \label{lumvv}
\end{equation}
Comparing that with equation~\eqref{lumjac}, we see that the van Vleck determinant is related to the Jacobi determinant via
\begin{equation}
\det J = (\lambda_o - \lambda_s)^2 \Delta_{vV}^{-1}.
\end{equation}

\subsection{Example: $k=0$ FLRW universe (without peculiar velocities)}
First we calculate the $d_L(z)$ relation in a $k=0$ FLRW universe expressed in terms of  conformal time $\eta$
\begin{equation}
\mathrm{d} s^2 = a^2(\eta) \mathrm{d}\hat{s}^2 = a^2(\eta) (-\mathrm{d}\eta^2 + \mathrm{d}x^2 + \mathrm{d}y^2 + \mathrm{d}z^2).
\label{frw}
\end{equation}

In this section and in later sections we make use of the following relations between quantities evaluated in conformally related metrics $\mathrm{d} s^2 = f^2(\eta, \vec{x}) \; \mathrm{d}\hat{s}^2$   and for timelike and null geodesic tangent vectors: $U^{\mu} = \frac{\mathrm{d} x^{\mu}}{\mathrm{d} \tau}$ and $\ell^{\mu} = \frac{\mathrm{d} x^{\mu}}{\mathrm{d} \lambda}$. Thereby

\begin{equation}
\frac{\mathrm{d} \lambda}{\mathrm{d} \hat{\lambda}} = f^2;
\qquad
\ell^{\mu} = \frac{1}{f^2} \; \hat{\ell}^{\mu};
\end{equation}

\begin{equation}
\frac{\mathrm{d} \tau}{\mathrm{d} \hat{\tau}} = f; \;
\qquad
U^{\mu} = \frac{1}{f} \; \hat{U}^{\mu}.
\end{equation}
This allows us to derive an expression relating the redshifts in the two conformal metrics
\begin{equation}
1+z = \frac{(g_{\mu \nu} k^{\mu} U^{\nu})_s}{(g_{\mu \nu} k^{\mu} U^{\nu})_o} = \frac{f_o (\hat{g}_{\mu \nu} \hat{k}^{\mu} \hat{U}^{\nu})_s} {f_s (\hat{g}_{\mu \nu} \hat{k}^{\mu} \hat{U}^{\nu})_o} = \frac{f_o}{f_s}\;(1+\hat{z}).
\end{equation}
The Jacobi map scales as
\begin{equation}
J(O,S) = f_o \;\hat{J}(O,S), \label{conf1}
\end{equation}
and therefore
\begin{equation}
\det  J(O,S) = f_o^2 \;\det  \hat{J}(O,S). \label{conf2}
\end{equation}
Hence, we have for the luminosity distances in conformally related spacetimes
\begin{align}
d_L &= (1+z)\;|\det J(O,S)|^{\frac{1}{2}} \nonumber \\
&= \frac{f_o^2}{f_s} \; (1+\hat{z})\; |\det  \hat{J}(O,S)|^{\frac{1}{2}} \nonumber \\
&= \frac{f_o^2}{f_s} \; \hat{d}_L. \label{jorah}
\end{align}
We want to use \eqref{jorah} and therefore we first want to compute $d_L$ for the reference Minkowski spacetime $ \mathrm{d}\hat{s}^2 = (-\mathrm{d}\eta^2 + \mathrm{d}x^2 + \mathrm{d}y^2 + \mathrm{d}z^2)$.
For now we shall take the source and the observer to be at rest with respect to each other and with respect to the Hubble flow, so that there are no peculiar velocities and no Doppler shift. Therefore the 4-velocities of the observer and the source in synchronous coordinates are
\begin{equation}
(\hat{U}^{\mu})_o = (1, 0, 0, 0);
\qquad
(\hat{U}^{\mu})_s = (1, 0, 0, 0).
\end{equation}
Here $\hat{\ell}^{\mu}$ is a null vector: $\hat{\ell}^{\mu} \hat{\ell}^{\nu} \eta_{\mu \nu} = 0$, and is the tangent vector to an affinely parameterized  null geodesic: $\frac{\mathrm{d} \hat{\ell}^{\mu}}{\mathrm{d} \hat{\lambda}} = 0$. Therefore, up to a normalization constant it has the following form 
\begin{equation}
\hat{\ell}^{\mu} = (1, \vec{n}),
\end{equation}
where $\vec{n}$ is a unit vector. Therefore, the wave vector is given by
\begin{equation}
\hat{k}^{\mu} = \tilde{\omega} (1, \vec{n}).
\end{equation}
Since the Christoffel symbols and the Riemann tensor vanish in Minkowski the system of equations 
\eqref{de2} reduces to
\begin{equation}
\frac{\mathrm{d}(\delta \hat{x}^{\alpha})}{\mathrm{d} \hat{\lambda}} = \delta \hat{\theta}^{\alpha}; 
\qquad
\frac{\mathrm{d}(\delta \hat{\theta}^{\alpha})}{\mathrm{d} \hat{\lambda}} = 0; \label{mink2}
\end{equation}
with the initial conditions \eqref{ic2}. It is easy to solve this system. The solution is
\begin{equation}
\delta \hat{\theta}^{\alpha} = \delta \hat{\theta}^{\alpha}_s;
\qquad
\delta \hat{x}^{\alpha}_o = (\hat{\lambda}_o - \hat{\lambda}_s) \delta \hat{\theta}^{\alpha}_s.
\end{equation}
Therefore the unprojected Jacobi map is simply
\begin{equation}
\hat{\mathcal{J}}^{\alpha}_{\beta} = (\hat{\lambda}_o - \hat{\lambda}_s)\, \delta^{\alpha}_{\beta}.
\label{unproj}
\end{equation}
We also have that:
\begin{equation}
\vec{n}_o = \vec{n}_s = \vec{n};
\qquad
\hat{P}_o = \hat{P}_s =: \hat{P}.
\end{equation}
The projected Jacobi map is
\begin{equation}
\hat{J}^{\mu}_{\nu} = \hat{P}^{\mu}_{\alpha} \hat{\mathcal{J}}^{\alpha}_{\beta} \hat{P}^{\beta}_{\nu}.
\end{equation}
It is easy to check that
\begin{equation}
\hat{J}^0_0 = \hat{J}^0_i = \hat{J}^i_0 = 0;
\qquad
\hat{J}^i_j = (\hat{\lambda}_o - \hat{\lambda}_s) (\delta^i_j - n^i n_j).
\end{equation}
The determinant is the product of the two non-vanishing eigenvalues
\begin{equation}
\det  \hat{J} = (\hat{\lambda}_o - \hat{\lambda}_s)^2.
\end{equation}
In Minkowski space, there is no gravitational redshift and we chose the observer and the source to be at rest with respect to each other and so there is no Doppler redshift. Hence, $\hat{z}=0$ and the luminosity distance becomes:
\begin{equation}
\hat{d}_L = (\hat{\lambda}_o - \hat{\lambda}_s).
\end{equation}
Now, using \eqref{jorah}, the luminosity distance in FLRW becomes
\begin{equation}
d_L = \frac{a_o^2}{a_s} \;\hat{d}_L = \frac{a_o^2}{a_s} \;(\hat{\lambda}_o - \hat{\lambda}_s) = \frac{a_o^2}{a_s} \; (\eta_o - \eta_s).
\end{equation}
One can cast this in a more familiar form by recognising that
\begin{equation}
d_m = a_o (\eta_o - \eta_s),
\end{equation}
and
\begin{equation}
1+z_c = \frac{a_o}{a_s},
\end{equation}
where $d_m$ is the metric distance and $z_c$ is the cosmological redshift in FLRW. Then the luminosity distance becomes
\begin{equation}
d_L = d_m (1+z_c).
\end{equation}

It is easy to check that the van Vleck approach also gives the correct formula for the luminosity distance in FLRW. In Minkowski space we have that
\begin{equation}
\hat{z} = 0;
\qquad
\hat{\Delta}_{vV} = 1.
\end{equation}
Hence, applying \eqref{lumvv}, we get the luminosity distance in Minkowski space
\begin{equation}
\hat{d}_L = \hat{\lambda}_o - \hat{\lambda}_s,
\end{equation}
and therefore the luminosity distance in FLRW is
\begin{equation}
d_L = \frac{a_o^2}{a_s}\;(\eta_o - \eta_s),
 \label{tom}
\end{equation}
It is useful to rewrite the luminosity distance as a perturbative power series in the redshift. Using the cosmographic expansion \eqref{E:cosmo} of section \ref{S:cosmo} we can write
\begin{align}
d_L(z) = \frac{a_o}{\mathcal{H}_o}\left[z - \frac{\mathcal{Q}_o}{2} z^2 
 + \frac{3\mathcal{Q}_o^2 + 3\mathcal{Q}_o - \mathcal{J}_o}{6} z^3 + O(z^4)\right].\label{henry2}
\end{align}

This is equivalent to \eqref{jackson} which was derived in~\cite{Visser}, except that now we are working with conformal time and consider terms only up to $O(z_s^3)$.
Note that
\begin{equation}
\frac{a_o}{\mathcal{H}_o} = \frac{a_o}{(da/d\eta)_o/a_o} = \frac{a_o}{(da/dt)_o}  = {1\over H_o},
\end{equation}
where $H_o$ is the usual Hubble parameter measured by the astronomers. That is, in any FLRW cosmology
\begin{align}
d_L(z) = \frac{1}{{H}_o}\left[z - \frac{\mathcal{Q}_o}{2} z^2 
 + \frac{3\mathcal{Q}_o^2 + 3\mathcal{Q}_o - \mathcal{J}_o}{6} z^3 + O(z^4)\right].\label{henry2b}
\end{align}
Furthermore, $\mathcal{Q}_o$ and $\mathcal{J}_o$ can be converted to $q_o$ and $j_o$ which gives \eqref{jackson} up to third order in $z$ as expected:
\begin{equation}
d_L(z) = \frac{1}{H_0} \Big[ z + \frac{1-q_0}{2}z^2 - \frac{1-q_0 - 3q_0^2 + j_0}{6} z^3 + O(z^4) \Big] \label{sam1}
\end{equation}
This result is with hindsight actually quite straightforward, and we can only justify the time spent on such an approach by now modifying and applying it in several non-trivial situations.

\subsection{Example: $k=0$ FLRW universe (with peculiar velocities)}
From the above we see that in a $k=0$ FLRW universe without peculiar velocities
\begin{equation}
d_L = \frac{a_o}{\mathcal{H}_o} (1+z) P(z),
\end{equation}
where $P(z)$ is the specific polynomial
\begin{align}
P(z) =  z - \frac{2+\mathcal{Q}_o}{2} z^2 
 +\frac{3\mathcal{Q}_o^2 + 6\mathcal{Q}_o + 6 - \mathcal{J}_o}{6} z^3 + O(z^4).
\end{align}
If one now adds peculiar velocities, then the only change is that
\begin{equation}
d_L = \frac{a_o}{\mathcal{H}_o} (1+z) P(z_c),
\end{equation}
where $z_c$ is the cosmological contribution to the total redshift $z$, and in terms of the Doppler contribution to the redshift we have
\begin{equation}
1+z= (1+z_c) (1+z_D).
\end{equation}
Then, assuming that peculiar velocities, and hence $z_D$, are small, we have
\begin{equation}
z_c = {1+z\over 1+z_D}-1 \approx z  - (1+z)z_D + O(z_D^2).
\end{equation}
Therefore
\begin{equation}
d_L = \frac{a_o}{\mathcal{H}_o} (1+z) P\left(z  - (1+z)z_D + O(z_D^2)\right),
\end{equation}
implying
\begin{equation}
d_L = \frac{a_o}{\mathcal{H}_o} \left\{ (1+z) P(z)   - (1+z)^2 P'(z) z_D + O(z_D^2)\right\}.
\end{equation}
This gives an explicit formula for estimating the potential effect of peculiar velocities on luminosity distance. 
The fractional size of the effect is easily seen to be
\begin{equation}
{\Delta d_L\over d_L} =   -  (1+z)  {P'(z)\over P(z)} z_D + O(z_D^2).
\end{equation}
Evaluating explicitly the polynomial $P(z)$ to $O(z^3)$, we can find an expression for $d_L$ to $O(z^2)$ and $O(z_D)$
\begin{align}
d_L &= \frac{a_o}{\mathcal{H}_o} \bigg[- z_D + \Big(1 + \mathcal{Q}_o z_D \Big) z 
\nonumber \\ 
&\qquad
- \bigg( \frac{\mathcal{Q}_o}{2}+ \frac{3\mathcal{Q}^2_o + 2\mathcal{Q}_o - \mathcal{J}_o}{2}z_D \bigg) z^2 + O(z^3) + O(z_D^2) \bigg].
\end{align}
As a further application we might consider a situation where on average the peculiar Doppler shifts are zero: $\langle z_D\rangle =0$.
Then on average
\begin{align}
\langle d_L \rangle &= \frac{a_o}{\mathcal{H}_o}
 \bigg[ z -\frac{\mathcal{Q}_o}{2} z^2 + O(z^3) + O(z_D^2) \bigg],
\end{align}
and so 
\begin{align}
d_L - \langle d_L\rangle &= -\frac{a_o \, z_D}{\mathcal{H}_o} \bigg[1 - \mathcal{Q}_o  z 
+ \bigg(\frac{3\mathcal{Q}^2_o + 2\mathcal{Q}_o - \mathcal{J}_o}{2} \bigg) z^2
+ O(z^3) + O(z_D) \bigg]. \label{PC1}
\end{align}
This could be used, in principle, to estimate peculiar Doppler redshifts $z_D$ (and so peculiar velocities) at various values of total redshift $z$. This would be done by first neglecting peculiar Doppler redshifts to naively fit $d_L(z)$ to the supernova data, thereby determining the cosmographic coefficients, and then binning the supernovae into small redshift bins to observationally determine $d_L - \langle d_L\rangle$. 
It will now be interesting to extend this perturbative analysis beyond simple FLRW universes.

\section{Introducing linear perturbations}\label{S:linear}
Now we look at a linearly perturbed FLRW metric with 2 scalar modes in the Newtonian gauge
\begin{equation}
\mathrm{d} s^2 = a^2(\eta) \Big[-(1+2\Psi(\vec{x}, \eta))\mathrm{d} \eta^2 + (1+2\Phi(\vec{x}, \eta))\delta_{ij} \mathrm{d}x^i \mathrm{d}x^j\Big],
\label{full}
\end{equation}
where $\Psi$ and $\Phi$ are the so called Bardeen potentials. From now on all quantities are expressed to first order in terms of the Bardeen potentials. To first order the metric \eqref{full} can be cast in the form
\begin{equation}
\mathrm{d} s^2 = f^2(\eta, \vec{x}) \Big[-(1+2\xi)\mathrm{d}\eta^2 +  \delta_{ij} \mathrm{d}x^i \mathrm{d}x^j\Big],
\end{equation}
where the overall conformal factor is
\begin{equation}
f(\eta, \vec{x}) = a(\eta) (1+2\Phi)^{\frac{1}{2}} \approx a(\eta) (1+\Phi),
\end{equation}
and
\begin{equation}
\xi = \Psi - \Phi.
\end{equation}

\subsection{Calculating the redshift}
Now look at the simplified one-mode metric
\begin{equation}
\mathrm{d}\hat{s}^2 = -(1+2\xi)\mathrm{d}\eta^2 +  \delta_{ij} \mathrm{d}x^i \mathrm{d}x^j,
\label{metr}
\end{equation}
which is simply background Minkowski space plus a perturbation:  $\hat{g}_{\mu \nu} = \eta_{\mu \nu} + h_{\mu \nu}$.
We require that the 4-velocities of the source and the observer are normalized $\hat{U}^{\mu} \hat{U}^{\nu} \hat{g}_{\mu \nu} = -1$ and we again first consider the case of zero peculiar velocities. This implies that to first order
\begin{equation}
\hat{U}_s^{\mu} = \hat{U}_o^{\mu} = (1-\xi, \vec{0})
\end{equation}
The source emits light which travels on null geodesics with wave vector $\hat{k}^{\mu}_s$. The emission frequency is given by
\begin{align}
\hat{\omega}_s &:= -\hat{g}_{\mu \nu} \hat{k}_s^{\mu} \hat{U}_s^{\nu} 
= -\tilde{\omega} \hat{g}_{\mu \nu} \hat{\ell}_s^{\mu} \hat{U}_s^{\nu} 
= \tilde{\omega} (1+\xi_s),
 \label{em}
\end{align}
where we use the fact that locally at the source spacetime is approximately flat. So $\hat{\ell}_s^{\mu} \approx \hat{\bar{\ell}}_s^{\mu} = (1, \vec{n})$ where the bar here and thereafter will denote the background value of a given object.
The observed frequency is similarly given by
\begin{equation}
\hat{\omega}_o = \tilde{\omega} \,\hat{\ell}_o^0 \,(1+\xi_o).
\end{equation}
Then the redshift is given by
\begin{equation}
1+\hat{z} = \frac{\hat{\omega}_s}{\hat{\omega}_o} = \frac{1+\xi_s}{\hat{\ell}_o^0 (1+\xi_o)} = (\hat{\ell}_o^0)^{-1} (1+\xi_s - \xi_o).
\end{equation}
In order to calculate this redshift, we need to relate the tangent vector of the light ray at the position of the observer $\hat{\ell}_o^{\mu}$ to the tangent vector at the source $\hat{\ell}_s^{\mu} \approx (1, \vec{n})$. 
This can be done via the geodesic equation
\begin{equation}
\frac{\mathrm{d} \hat{\ell}^{\mu}}{\mathrm{d} \hat{\lambda}} = - \hat{\Gamma}^{\mu}_{\rho \sigma} \hat{\ell}^{\rho} \hat{\ell}^{\sigma},\label{geod}
\end{equation}
which to first order becomes
\begin{equation}
\frac{\mathrm{d} \hat{\ell}^{(1)\mu}}{\mathrm{d} \hat{\lambda}} = - \hat{\Gamma}^{\mu}_{\rho \sigma} \hat{\bar{\ell}}^{\rho} \hat{\bar{\ell}}^{\sigma},
\end{equation}
where the background connection vanishes: $\hat{\bar{\Gamma}}^{\mu}_{\rho \sigma} = 0$ because the background space is Minkowski. The Christoffel symbols can be easily calculated from the metric \eqref{metr}
\begin{equation}
\hat{\Gamma}^{0}_{00} = \xi_{,\eta};
\end{equation}
\begin{equation}
\hat{\Gamma}^{0}_{0i} = \hat{\Gamma}^{0}_{i0} = \hat{\Gamma}^{i}_{00} = \xi_{,i} ;
\end{equation}
\begin{equation}
\hat{\Gamma}^{0}_{ij} = \hat{\Gamma}^{k}_{0i} = \hat{\Gamma}^{k}_{i0} = \hat{\Gamma}^{k}_{ij} = 0;
\end{equation}
Hence, the solution of the geodesic equation is given by
\begin{equation}
\hat{\ell}_o^{(1)0} - \hat{\ell}_s^{(1)0} = - \int^{\hat{\lambda}_o}_{\hat{\lambda}_s} (\xi_{,\eta} + 2\vec{\nabla} \xi . \vec{n}) \mathrm{d}\hat{\lambda}; \label{sol0}
\end{equation}
\begin{equation}
\hat{\ell}_o^{(1)i} -\hat{\ell}_s^{(1)i} = - \int^{\hat{\lambda}_o}_{\hat{\lambda}_s} \xi_{,i} \mathrm{d}\hat{\lambda}.
\end{equation}
Equation \eqref{sol0}, and the fact that $\hat{\ell}_s^{\mu} \approx (1, \vec{n})$, together imply that
\begin{equation}
\hat{\ell}_o^0 = 1 - \int^{\hat{\lambda}_o}_{\hat{\lambda}_s} (\xi_{,\eta} + 2\vec{\nabla} \xi \cdot \vec{n}) \mathrm{d} \hat{\lambda}.
\label{winston}
\end{equation}
Therefore, the redshift to first order becomes
\begin{equation}
1+\hat{z} = 1 - (\xi_o - \xi_s) + \int^{\hat{\lambda}_o}_{\hat{\lambda}_s} (\xi_{,\eta} + 2\vec{\nabla} \xi \cdot \vec{n}) \mathrm{d}\hat{\lambda}.
\end{equation}
We can put that in a more useful form by changing variables from the affine parameter $\hat{\lambda}$ to the conformal time $\eta$. Using  \eqref{winston} we have that to first order
\begin{equation}
\mathrm{d} \hat{\lambda} = \mathrm{d} \eta \left[1 + \int^{\hat{\lambda}}_{\hat{\lambda}_s} (\xi_{,\eta} + 2\vec{\nabla}\xi \cdot \hat{n}) \mathrm{d}\hat{\lambda}^{\prime}\right].
\end{equation}
We also use that to first order
\begin{equation}
\xi_{,\eta} = \frac{\mathrm{d}\xi}{\mathrm{d}\eta} - \vec{\nabla}\xi \cdot \vec{n}.
\end{equation}
This then gives us the following expression for the redshift
\begin{equation}
1+\hat{z} = 1 + \int^{\eta_o}_{\eta_s} \vec{\nabla} \xi \cdot \vec{n} \mathrm{d} \eta.
 \label{george}
\end{equation}
or, equivalently,
\begin{equation}
1+\hat{z} = 1 + \xi_o - \xi_s - \int^{\eta_o}_{\eta_s} \xi_{,\eta} \mathrm{d} \eta.
 \label{george2}
\end{equation}
Now we can find the redshift in the full perturbed FLRW metric
\begin{equation}
1+z = \frac{a_o}{a_s} \left(1 + \Phi_o - \Phi_s + \int^{\eta_o}_{\eta_s} \vec{\nabla} \xi \cdot \vec{n} \mathrm{d} \eta\right).
 \label{redshift}
\end{equation}
or, equivalently,
\begin{equation}
1+z = \frac{a_o}{a_s} \Big( 1 + \Psi_o - \Psi_s - \int^{\eta_o}_{\eta_s} \xi_{,\eta} \mathrm{d}\eta \Big)
\end{equation}
The redshift is a product of different contributions
\begin{equation}
1+z = (1+z_{c})(1+z_{gr})(1+z_{ISW}). \label{indiana}
\end{equation}
Here 
\begin{equation}
1+z_c = \frac{a_o}{a_s},
\end{equation}
is the cosmological redshift due to the overall expansion of the universe, and
\begin{equation}
1+z_{gr} = \sqrt{\frac{1+2\Psi_o}{1+2\Psi_s}} \approx 1+\Psi_o - \Psi_s,
\end{equation}
is the gravitational redshift due to the potential wells of the source and the observers. 
Finally 
\begin{equation}
1+z_{ISW} = 1 - \int^{\eta_o}_{\eta_s} \xi_{,\eta} \mathrm{d}\eta
\;  = 1 - \int_{t_s}^{t_o}  \xi_{,t} dt
\end{equation}
is the gravitational redshift caused by changing potential wells along the path of the light --- an integrated Sachs--Wolfe effect~\cite{Sachs}. Eqn. \eqref{indiana} gives the total redshift without the Doppler redshift arising due to the peculiar velocities of the source and the observer. It is trivial to include the Doppler redshift in the analysis - \eqref{indiana} is modified to
\begin{equation}
1+z = (1+z_{D})(1+z_{c})(1+z_{gr})(1+z_{ISW}),
\end{equation}
where the Doppler contribution to the redshift is
\begin{equation}
1+z_D = \frac{\gamma_s (1-\vec{v}_s . \vec{n})}{\gamma_o (1-\vec{v}_o . \vec{n})},
\end{equation}
and where $\gamma = (1 - |\vec{v}|^2)^{-\frac{1}{2}}$ and $\vec{v}_s$, $\vec{v}_o$ are the peculiar velocities of the source and the observer. 

We can also adapt this redshift calculation to determine the total lapse in affine parameter in terms of the total lapse in conformal time.
From the above, the relationship between affine parameter and conformal time is
\begin{eqnarray}
\mathrm{d} \hat{\lambda} &=& 
\mathrm{d} \eta \left[1 + \int^{\hat{\lambda}}_{\hat{\lambda}_s} (\xi_{,\eta} + 2\vec{\nabla}\xi \cdot \hat{n}) \mathrm{d}\hat{\lambda}^{\prime}\right] \nonumber
\\
&=&
\mathrm{d} \eta \left[1 +  2(\xi - \xi_s) - \int^{\eta}_{\eta_s} \xi_{,\eta^{\prime}} \mathrm{d} \eta^{\prime} \right] \nonumber
\\
&=& 
\mathrm{d} \eta \left[1 +  2(\xi - \xi_s) + z_{ISW}(\eta_s) - z_{ISW}(\eta)\right].
\end{eqnarray}
where
\begin{equation}
z_{ISW}(\eta) = - \int^{\eta_o}_{\eta} \xi_{,\eta^{\prime}} \mathrm{d} \eta^{\prime}.
\end{equation}
Integrating
\begin{equation}
\hat\lambda_o-\hat\lambda_s  = 
(\eta_o-\eta_s) \; \left[1+2( \langle\xi\rangle - \xi_s) +z_{ISW}- \langle  z_{ISW} \rangle \right]. \label{hahaha}
\end{equation}
Here $ \langle\xi\rangle$ and $\langle  z_{ISW} \rangle$ are simply averages along the line of sight:
\begin{equation}
\langle \xi \rangle := \frac{1}{\eta_o - \eta_s} \int^{\eta_o}_{\eta_s} \xi \mathrm{d} \eta;
\end{equation}
\begin{equation}
\langle z_{ISW} \rangle := \frac{1}{\eta_o - \eta_s} \int^{\eta_o}_{\eta_s} z_{ISW} (\eta) \mathrm{d} \eta.
\end{equation}
While  $ \langle\xi\rangle$ and $\langle  z_{ISW} \rangle$, (and $\xi_s$ and $z_{ISW}$ for that matter), might be difficult to measure, they do at least have clear physical interpretations.

\subsection{The Jacobi and van Vleck determinants}
The Jacobi map and Jacobi determinant can be calculated using the formalism developed in section \ref{Jmd}. We present here the final result for the Jacobi determinant and defer the full calculation to Appendix \ref{A:2}. The Jacobi determinant in the unphysical metric \eqref{metr} is given by:
\begin{align}
(\det \hat{J})^{\frac{1}{2}} &= (\hat{\lambda}_o - \hat{\lambda}_s)   \left\{ 1 
-\frac{1}{2} {1\over \hat{\lambda}_o - \hat{\lambda}_s} \int^{\hat{\lambda}_o}_{\hat{\lambda}_s} (\hat{\lambda}_o - \hat{\lambda}) (\nabla^2 \xi - n^i n^j \xi_{,ij}) (\hat{\lambda} - \hat{\lambda}_s) \mathrm{d}\hat{\lambda}\right\}.
 \label{jacdet}
\end{align}
If the Jacobi and the van Vleck approaches are equivalent, as was non-perturbatively demonstrated in~\cite{Non-perturbative},  we must have that
\begin{equation}
(\det  \hat{J})^{\frac{1}{2}} = \hat{\Delta}_{vV}^{-\frac{1}{2}} \; (\hat{\lambda}_o - \hat{\lambda}_s).
 \label{tim}
\end{equation}
We will now show that this is indeed the case.

In the weak field limit the van Vleck determinant is approximated by~\cite{vanVleck,Visser:1995,Visser:1993}
\begin{align}
\hat{\Delta}_{vV} &\approx \mathrm{exp}\left[ \frac{1}{\hat{\lambda}_o - \hat{\lambda}_s} \int^{\hat{\lambda}_o}_{\hat{\lambda}_s} (\hat{\lambda}_o - \hat{\lambda})(\hat{R}_{\mu \nu} \hat{\ell}^{\mu} \hat{\ell}^{\nu}) (\hat{\lambda} - \hat{\lambda}_s) \mathrm{d} \hat{\lambda}\right]; \nonumber \\
&\approx 1 + \left[ \frac{1}{\hat{\lambda}_o - \hat{\lambda}_s} \int^{\hat{\lambda}_o}_{\hat{\lambda}_s} (\hat{\lambda}_o - \hat{\lambda})(\hat{R}_{\mu \nu} \hat{\ell}^{\mu} \hat{\ell}^{\nu}) (\hat{\lambda} - \hat{\lambda}_s) \mathrm{d} \hat{\lambda}\right].
 \label{sam}
\end{align}
The components of the Ricci tensor to first order are
\begin{equation}
\hat{R}_{00} = \nabla^2 \xi; 
\qquad
\hat{R}_{0i} = 0;
\qquad
\hat{R}_{ij} = - \xi_{,ij}.
\end{equation}
Since $\hat{\bar{R}}_{\mu \nu}=0$, only the term $\hat{R}^{(1)}_{\mu \nu} \hat{\bar{\ell}}^{\mu} \hat{\bar{\ell}}^{\nu}$ will contribute to first order in the expression \eqref{sam}.
We have
\begin{equation}
\hat{R}^{(1)}_{\mu \nu} \, \hat{\bar{\ell}}^{\mu} \hat{\bar{\ell}}^{\nu} = (\nabla^2 \xi - n^i n^j \xi_{,ij}),
\end{equation}
and therefore
\begin{equation}
\hat{\Delta}_{vV} = 1 + \frac{1}{\hat{\lambda}_o - \hat{\lambda}_s} \int^{\hat{\lambda}_o}_{\hat{\lambda}_s} (\hat{\lambda}_o - \hat{\lambda}) (\nabla^2 \xi - n^i n^j \xi_{,ij}) (\hat{\lambda} - \hat{\lambda}_s) \mathrm{d} \hat{\lambda}.
\end{equation}
Hence,
\begin{equation}
\hat{\Delta}_{vV}^{-\frac{1}{2}} = 1 - \frac{1}{\hat{\lambda}_o - \hat{\lambda}_s} \frac{1}{2} \int^{\hat{\lambda}_o}_{\hat{\lambda}_s} (\hat{\lambda}_o - \hat{\lambda}) (\nabla^2 \xi - n^i n^j \xi_{,ij}) (\hat{\lambda} - \hat{\lambda}_s) \mathrm{d} \hat{\lambda}.
\end{equation}
We see that \eqref{tim} is satisfied and so the two approaches are equivalent.

\subsection{The luminosity distance in perturbed FLRW}
Now we finish the calculation of the luminosity distance in perturbed FLRW. We can express the Jabobi determinant \eqref{jacdet} in terms of conformal time by using the fact that
\begin{equation}
\frac{\mathrm{d}\eta}{\mathrm{d} \hat{\lambda}} = 1 - \int^{\hat{\lambda}}_{\hat{\lambda}_s} (\xi_{,\eta} + 2 \vec{\nabla}\xi \cdot \vec{n}) \mathrm{d} \hat{\lambda}^{\prime},
\end{equation}
and hence to linear order
\begin{equation}
\mathrm{d} \hat{\lambda} = \mathrm{d} \eta \left(1 + \int^{\eta}_{\eta_s} (\xi_{,\eta} + 2 \vec{\nabla}\xi \cdot \vec{n}) \mathrm{d} \eta^{\prime}\right).
\end{equation}
The resulting expression for the Jacobi determinant is:
\begin{align}
(\det \hat{J})^{\frac{1}{2}} &= (\eta_o - \eta_s) + \int^{\eta_o}_{\eta_s} \xi \mathrm{d}\eta + \int^{\eta_o}_{\eta_s} (\eta_o - \eta) (\vec{\nabla}\xi \cdot \vec{n}) \mathrm{d}\eta - \xi_s (\eta_o - \eta_s) \\ \nonumber
& -\frac{1}{2} \int^{\eta_o}_{\eta_s} (\eta_o - \eta) (\nabla^2 \xi - n^i n^j \xi_{,ij}) (\eta - \eta_s) \mathrm{d}\eta,
\end{align}
where again we have replaced a double integral by a single integral.  
Hence, the luminosity distance in two-mode perturbed ($\Phi$, $\xi$) FLRW cosmology is given by
\begin{align}
d_L(\eta_s, \eta_o, \vec{n}) &= \frac{f_o^2}{f_s}\; \hat{d}_L \\
&= \frac{f_o^2}{f_s} \;(\det \hat{J})^{\frac{1}{2}} \;(1+\hat{z}) \label{tender}\\
&= \frac{a_o^2}{a_s}\;\Big[(\eta_o - \eta_s) + 2\Phi_o (\eta_o - \eta_s) - \Psi_s (\eta_o - \eta_s) \nonumber \\
& \qquad+ (\eta_o - \eta_s) \int^{\eta_o}_{\eta_s} \vec{\nabla} \xi \cdot \vec{n} \mathrm{d}\eta + \int^{\eta_o}_{\eta_s} \xi \mathrm{d} \eta + \int^{\eta_o}_{\eta_s} (\eta_o - \eta) (\vec{\nabla}\xi \cdot \vec{n}) \mathrm{d}\eta \nonumber \\
&\qquad -\frac{1}{2} \int^{\eta_o}_{\eta_s} (\eta_o - \eta) (\nabla^2 \xi - n^i n^j \xi_{,ij}) (\eta - \eta_s) \mathrm{d}\eta \Big].
\label{lumdis}
\end{align}
This formula shows the dependance of the luminosity distance measured by an observer ${O}$ as a function of the conformal time of the source $\eta_s$ in a given direction $\vec{n}$. 
At this stage, this expression is somewhat formal, and mainly useful as a starting point for further detailed model-building.
We shall present a particularly simple toy model in the next section, but for now will try to re-cast this expression (to the extent possible) in terms of various contributions to the redshift. 
For instance, by recognizing that $d_{L, FLRW} = \frac{a_o^2}{a_s} (\eta_o - \eta_s)$ is the luminosity distance in FLRW without peculiar velocities, one can write
\begin{eqnarray}
d_L(\eta_s, \eta_o, \vec{n})  &=&  d_{L,FLRW}(z_c) \bigg[ 1+2\Phi_o-\Psi_s 
  + \int^{\eta_o}_{\eta_s} \vec{\nabla} \xi \cdot \vec{n} \mathrm{d}\eta 
\nonumber\\
&&+ {1\over\eta_o-\eta_s}\int^{\eta_o}_{\eta_s} \xi \mathrm{d} \eta +{1\over\eta_o-\eta_s} \int^{\eta_o}_{\eta_s} (\eta_o - \eta) (\vec{\nabla}\xi \cdot \vec{n}) \mathrm{d}\eta 
\nonumber \\
&&\qquad -\frac{1}{2} {1\over\eta_o-\eta_s}\int^{\eta_o}_{\eta_s} (\eta_o - \eta) (\nabla^2 \xi - n^i n^j \xi_{,ij}) (\eta - \eta_s) \mathrm{d}\eta \bigg].
\label{lumdis2}
\end{eqnarray}
There are several other ways of usefully repackaging the luminosity distance in the two-mode perturbed ($\Phi$, $\xi$) FLRW cosmology we are considering. For instance, using \eqref{hahaha}, we have that
\begin{equation}
(\det \hat{J})^{\frac{1}{2}} = (\eta_o-\eta_s) \; \left[1+2( \langle\xi\rangle - \xi_s) +z_{ISW}- \langle  z_{ISW} \rangle \right] \hat{\Delta}_{vV}^{- \frac{1}{2}}
\end{equation}
and substituting that inside \eqref{tender} we obtain
\begin{eqnarray}
d_L &=& d_{L,FLRW}(z_c) \,(1+\Phi_o)\,(1+z_{gr})\,(1+z_{ISW})\,[1+2(\langle \xi\rangle-\xi_s)+z_{ISW}-\langle z_{ISW}\rangle]
\nonumber\\
&& \qquad \times \left\{ 1   -\frac{1}{2} {1\over\eta_o-\eta_s}\int^{\eta_o}_{\eta_s} (\eta_o - \eta) (\nabla^2 \xi - n^i n^j \xi_{,ij}) (\eta - \eta_s) \mathrm{d}\eta \right\}.
\label{lumdis3}
\end{eqnarray}
The $(1+\Phi_o)$ factor is relatively uninteresting, since it only depends on what is happening at the observer, it is common to all observations --- at worst it is a rescaling to marginalize over.
These various ways of looking at the luminosity distance, we do feel, give us a somewhat better handle on the fundamental physics.
Equations \eqref{lumdis2} and \eqref{lumdis3} are now manifestly of the form
\begin{equation}
d_L(\eta_s, \eta_o, \vec{n})  =  d_{L,FLRW}(z_c) \times \left\{ 1 + \hbox{(perturbatively small)} \right\}.
\end{equation}

\section{Simple toy model: \\Scalar mode perturbation sinusoidally varying with time}\label{S:toy}

We now consider a simple toy model where the Bardeen potentials depend sinusoidally on conformal time and are independent of space
\begin{equation}
\Psi = -\Phi = \epsilon\; \sin(\kappa \eta).
\end{equation}
where $\epsilon$ and $\kappa$ are constants and $\epsilon$ is perturbatively small. Initially we shall neglect peculiar velocities, but subsequently show how to put them back in. We choose this particular toy model because it is tractable, and because it serves to illustrate the basic principles behind generalising the cosmographic approach to an inhomogeneous universe. Obviously, in order to analyse the real universe, one would need to consider more sophisticated models.

\subsection{Toy model without peculiar velocities}

Equations \eqref{lumdis} and \eqref{redshift} become
\begin{equation}
d_L = \frac{a_o^2}{a_s} \Big[ \Delta \eta + \epsilon \Big( - 2 \sin (\kappa \eta_o) \Delta \eta - \sin (\kappa \eta_s) \Delta \eta - 2 \frac{\cos (\kappa \eta_o)}{\kappa} + 2 \frac{\cos (\kappa \eta_s)}{\kappa} \Big) \Big];\,
\label{joshua}
\end{equation}
and
\begin{equation}
1+z_s = \frac{a_o}{a_s} \Big[1 + \epsilon \Big(- \sin (\kappa \eta_o) + \sin(\kappa \eta_s) \Big) \Big].
\end{equation}
Now we derive a cosmographic series for $d_L$ in terms of $z$. The coefficients to leading order are expected to be the same as in \eqref{henry2} plus corrections of order $\epsilon$. The cosmographic parameters are defined in the same way as before -- equations \eqref{e:H_eta}, and we make use of the following relation, valid for any conformal time $\eta$,
\begin{equation}
1+z(\eta) = \frac{a_o}{a(\eta)} \Big[1 + \epsilon \Big(- \sin (\kappa \eta_o) + \sin(\kappa \eta) \Big) \Big].\label{actor}
\end{equation}
Expanding $a(\eta)$ and $\sin(\eta)$ as a series in terms of $(\eta - \eta_o)$ inside \eqref{actor}, we obtain a series for $z(\eta)$ in terms of $(\eta - \eta_o)$
\begin{align}
z(\eta) &= \bigg[- \mathcal{H}_o + \epsilon\, \Big( \kappa \cos (\kappa \eta_o)\Big) \bigg](\eta - \eta_o) \nonumber \\
&+ \left[\mathcal{H}_o^2 \Big(\frac{2+\mathcal{Q}_o}{2}\Big) + \epsilon \Big(- \kappa \cos(\kappa \eta_o) \mathcal{H}_o - \kappa^2 \frac{\sin(\kappa \eta_o)}{2}\Big) \right](\eta - \eta_o)^2 \nonumber \\
&+ \bigg[- \mathcal{H}_o^3 \Big(\frac{\mathcal{J}_o + 6\mathcal{Q}_o + 6}{6}\Big) + \epsilon \Big( \kappa \cos(\kappa \eta_o) \mathcal{H}_o^2 (\frac{2+\mathcal{Q}_o}{2}) \nonumber \\
&+ \kappa^2 \frac{\sin(\kappa \eta_o)}{2} \mathcal{H}_o - \kappa^3 \frac{\cos(\kappa \eta_o)}{6} \Big) \bigg] (\eta-\eta_o)^3 + O(\eta - \eta_o)^4.
\end{align}
Reverting this series, we find
\begin{equation}
\eta - \eta_o = A_1\,z + A_2\,z^2 + A_3\,z^3 + O(z^4),
\end{equation}
where
\begin{equation}
A_1 = -\frac{1}{\mathcal{H}_o} + \epsilon \bigg[ -\kappa \frac{\mathrm{cos}(\kappa \eta_o)}{\mathcal{H}_o^2} \bigg];
\end{equation}

\begin{align}
A_2 &= \frac{1}{\mathcal{H}_o} \Big( \frac{2+\mathcal{Q}_o}{2} \Big) \nonumber \\
&+ \epsilon \bigg[ -\kappa \frac{\mathrm{cos}(\kappa \eta_o)}{\mathcal{H}_o^2} - \kappa^2 \frac{\mathrm{sin}(\eta_o)}{2\mathcal{H}_o^3} + \kappa \frac{3 \mathrm{cos}(\eta_o)}{\mathcal{H}_o^2} \Big(\frac{2+\mathcal{Q}_o}{2} \Big) \bigg];
\end{align}

\begin{align}
A_3 &= -\frac{1}{\mathcal{H}_o}\left(\frac{6+3\mathcal{Q}_o^2 + 6\mathcal{Q}_o - \mathcal{J}_o}{6}\right) \nonumber \\
&+ \epsilon \frac{1}{\mathcal{H}_o^2} \bigg[ \kappa \cos(\kappa \eta_o) \bigg( \frac{-6-9\mathcal{Q}_o + 5\mathcal{Q}_o^2 - 2\mathcal{J}_o^2}{2} \bigg) \nonumber \\
&+ \kappa^3 \frac{\cos(\kappa \eta_o)}{6 \mathcal{H}_o^2} + \kappa \frac{\sin(\kappa \eta_o)}{\mathcal{H}_o} \bigg( \frac{3+2\mathcal{Q}_o}{2} \bigg) \bigg]
\end{align}
We also have
\begin{equation}
\Delta \eta = \eta_o - \eta_s = -A_1z_s - A_2 z_s^2 - A_3 z_s^3 + O(z_s^4).
\end{equation}
This allows us to expand $\sin(\eta_s)$, $\cos(\eta_s)$ and $\frac{a_o}{a_s}$ as functions of $z_s$. We find
\begin{align}
\sin(\kappa \eta_s) &= \sin(\kappa \eta_o) + \bigg[ \kappa \cos(\kappa \eta_o) A_1 \bigg] z_s \nonumber \\
&+\left[\kappa \cos(\kappa \eta_o)A_2 - \kappa^2\frac{\sin(\kappa \eta_o)A_1^2}{2}\right]z_s^2 \nonumber \\
&+\left[\kappa \cos(\kappa \eta_o) A_3 - \kappa^2 \sin(\kappa \eta_o) A_1 A_2 - \kappa^3 \frac{\cos(\kappa \eta_o)}{6} A_1^3\right]z_s^3 + O(z_s^4);
\end{align}
while
\begin{align}
\cos(\kappa \eta_s) &= \cos(\kappa \eta_o) + \bigg[-\kappa \sin(\kappa \eta_o) A_1\bigg] z_s \nonumber \\
&+\left[-\kappa \sin(\kappa \eta_o)A_2 - \frac{\kappa^2 \cos(\kappa \eta_o)A_1^2}{2}\right]z_s^2 \nonumber \\
&+\left[-\kappa \sin(\kappa \eta_o) A_3 - \kappa^2 \cos(\kappa \eta_o) A_1 A_2 + \kappa^3\frac{\sin(\kappa \eta_o)}{6} A_1^3\right]z_s^3 + O(z_s^4);
\end{align}
and
\begin{align}
\frac{a_o}{a_s} &= 1 + \bigg[1-\epsilon \kappa \cos(\kappa \eta_o) A_1\bigg] z_s \nonumber \\
&+\epsilon 
\left[- \kappa \cos(\kappa \eta_o)A_2 + \kappa^2 \frac{\sin(\kappa \eta_o)A_1^2}{2} - \kappa \cos(\kappa \eta_o) A_1\right]z_s^2 \nonumber \\
&-\epsilon \bigg[\kappa \cos(\kappa \eta_o) A_3 - \kappa^2 \sin(\kappa \eta_o) A_1 A_2 - \kappa^3 \frac{\cos(\kappa \eta_o)}{6} A_1^3 \nonumber \\
&+\kappa \cos(\kappa \eta_o)A_2 - \kappa^2 \frac{\sin(\kappa \eta_o)A_1^2}{2} \bigg]z_s^3 + O(z_s^4).
\end{align} 
Substituting everything inside equation \eqref{joshua} we obtain an expansion of the luminosity distance $d_L$ in terms of the redshift $z$
\begin{align}
\frac{d_L}{a_o} &= \left[\frac{1}{\mathcal{H}_o} + \epsilon \mathcal{X}\right] z_s 
+ \left[-\frac{\mathcal{Q}_o}{2\mathcal{H}_o} + \epsilon \mathcal{Y}\right] z_s^2 
\nonumber \\
&
+ \left[\frac{1}{\mathcal{H}_o} \bigg(\frac{3\mathcal{Q}_o^2 + 3\mathcal{Q}_o - \mathcal{J}_o}{6} \bigg) + \epsilon \mathcal{Z}\right] z_s^3 + O(z_s^4). 
\label{aragorn}
\end{align}
Here
\begin{equation}
\mathcal{X} := 2\frac{\sin(\kappa \eta_o)}{\kappa} - 2\frac{\cos(\kappa \eta_o)}{\kappa} - \frac{\sin(\kappa \eta_o)}{\mathcal{H}_o} + \kappa \frac{\cos(\kappa \eta_o)}{\mathcal{H}_o^2};
\end{equation}
\begin{equation}
\mathcal{Y} := \kappa \frac{6\cos(\kappa \eta_o)}{\mathcal{H}_o^2} + \kappa \frac{3\cos(\kappa \eta_o) \mathcal{Q}_o}{2\mathcal{H}_o^2} + \frac{\sin(\kappa \eta_o) \mathcal{Q}_o}{2 \mathcal{H}_o} + \kappa^2\frac{\sin(\kappa \eta_o)}{2\mathcal{H}_o^3};
\end{equation}
and
\begin{align}
\mathcal{Z} &:= \kappa \frac{2\cos(\kappa \eta_o)}{\mathcal{H}_o^2} - \kappa^2 \frac{2\cos(\kappa \eta_o)}{\mathcal{H}_o^3} - \kappa^2 \frac{\cos(\kappa \eta_o) \mathcal{Q}_o}{\mathcal{H}_o^3} - \kappa^3 \frac{\cos(\kappa \eta_o)}{6 \mathcal{H}_o^4} \nonumber \\
&+ \kappa \frac{\cos(\kappa \eta_o)}{\mathcal{H}_o^2} \left(\frac{36 + 15\mathcal{Q}_o^2 + 36\mathcal{Q}_o - 4\mathcal{J}_o}{6}\right) - \frac{4\sin(\kappa \eta_o)}{\mathcal{H}_o} - \kappa^2 \frac{\sin(\kappa \eta_o)}{3 \mathcal{H}_o^3} \nonumber \\
&- \frac{\sin(\kappa \eta_o) \mathcal{Q}_o}{2 \mathcal{H}_o} - \kappa^2 \frac{\sin(\kappa \eta_o) \mathcal{Q}_o}{\mathcal{H}_o^3} + \frac{2 \sin(\kappa \eta_o)}{\mathcal{H}_o}\left(\frac{6 + 3\mathcal{Q}_o^2 + 6\mathcal{Q}_o - \mathcal{J}_o}{6}\right).
\end{align}
This agrees to zeroth order in $\epsilon$ with equation \eqref{henry2}.

From the above we see that in our toy model (a sinusoidally perturbed $k=0$ FLRW universe) without peculiar velocities we have
\begin{equation}
d_L = \frac{a_o}{\mathcal{H}_o} (1+z) P(z),
\end{equation}
where $P(z)$ is the specific polynomial
\begin{align}
P(z) = &B_1\,z + B_2\,z^2 + B_3\,z^3 + O(z^4).
\end{align}
with
\begin{eqnarray}
B_1 &=& 1 + \epsilon \mathcal{H}_o \mathcal{X};
\\
B_2 &=& -\frac{\mathcal{Q}_o + 2}{2} + \epsilon \mathcal{H}_o (\mathcal{Y} - \mathcal{X});
\\
B_3 &=& \frac{3\mathcal{Q}_o^2 + 6\mathcal{Q}_o - \mathcal{J}_o + 6}{6} + \epsilon \mathcal{H}_o (\mathcal{Z} - \mathcal{Y} + \mathcal{X}).
\end{eqnarray}
The only thing that has changed with respect to standard FLRW is the coefficients of the polynomial.

\subsection{Toy model with peculiar velocities}

If one now adds peculiar velocities, then again the only change is that
\begin{equation}
d_L = \frac{a_o}{\mathcal{H}_o} (1+z) P(z_c),
\end{equation}
where $z_c$ is the cosmological contribution to the total redshift $z$. Now in terms of the redshift contributions due to peculiar velocity $z_p$ we again have
\begin{equation}
z_c = {1+z\over 1+z_D}-1 \approx z  - (1+z)z_D + O(z_D^2),
\end{equation}
again implying
\begin{equation}
d_L = \frac{a_o}{\mathcal{H}_o} \left\{ (1+z) P(z)   - (1+z)^2 P'(z) z_D + O(z_D^2)\right\},
\end{equation}
Within the context of this model universe, this gives an explicit formula for estimating the potential effect of peculiar velocities on the luminosity distance. 
Again evaluating explicitly the polynomial $P(z)$ to $O(z^3)$ allows us to express $d_L$ to $O(z^2)$ and $O(z_D)$
\begin{align}
d_L &= \frac{a_o}{\mathcal{H}_o} \bigg\{ - \bigg( 1 + \epsilon \mathcal{H}_o \mathcal{X} \bigg) z_D \nonumber \\
&+ \bigg[ 1 + \epsilon \mathcal{H}_o \mathcal{X} - \bigg( -\mathcal{Q}_o + \epsilon 2 \mathcal{H}_o \mathcal{Y} \bigg) z_D \bigg] z \nonumber \\
& + \bigg[ -\frac{\mathcal{Q}_o}{2} + \epsilon \mathcal{H}_o \mathcal{Y} - \bigg( \frac{3\mathcal{Q}_o^2 + 2\mathcal{Q}_o - \mathcal{J}_o}{2} + \epsilon \mathcal{H}_o (\mathcal{Y} + 3\mathcal{Z}) \bigg) z_D \bigg] z^2 \nonumber \\
&+ O(z^3) + O(z_D^2) \bigg\} .
\end{align}

We could proceed further for instance by assuming $\langle z_D\rangle=0$, (effectively temporarily ignoring peculiar Doppler shifts), and fitting 
\begin{align}
\langle d_L\rangle &= 
 \frac{a_o}{\mathcal{H}_o} \bigg\{ \bigg[ 1 + \epsilon \mathcal{H}_o \mathcal{X} \bigg] z 
+ \bigg[ -\frac{\mathcal{Q}_o}{2} + \epsilon \mathcal{H}_o \mathcal{Y} 
\bigg] z^2 
+ O(z^3) \bigg\}.
\end{align}
Then
\begin{align}
d_L-\langle d_L \rangle &= 
- \frac{a_o}{\mathcal{H}_o} z_D \bigg\{ \bigg( 1 + \epsilon \mathcal{H}_o \mathcal{X} \bigg)
+   \bigg( -\mathcal{Q}_o + \epsilon 2 \mathcal{H}_o \mathcal{Y} \bigg) z 
\nonumber \\
&\qquad + \bigg( \frac{3\mathcal{Q}_o^2 + 2\mathcal{Q}_o - \mathcal{J}_o}{2} + \epsilon \mathcal{H}_o (\mathcal{Y} + 3\mathcal{Z}) \bigg) z^2 \bigg\}
+ O(z^3) + O(z_D^2). \label{PC2}
\end{align}
So even in this sinusoidally perturbed FLRW model we see how we can use cosmographic techniques to estimate the size of the peculiar Doppler shifts.

\section{Summary and discussion}\label{S:summary}

In this paper we have derived a theoretical relation between the luminosity distance and the redshift of a standardizable candle in a linearly perturbed FLRW universe \eqref{full}. The relation is given by two equations \eqref{redshift} and \eqref{lumdis} 
\begin{equation}
1+z = \frac{a_o}{a_s}\left(1 + \Phi_o - \Phi_s + \int^{\eta_o}_{\eta_s} \vec{\nabla} \xi \cdot \hat{n} \mathrm{d} \eta\right)
=
(1+z_c)(1+z_{gr}) \left(1 + z_{ISW} \right),
 \label{redshift2}
\end{equation}
and
\begin{eqnarray}
d_L &=& d_{L,FLRW}(z_c) \,(1+\Phi_o)\,(1+z_{gr})\,(1+z_{ISW})\,[1+2(\langle \xi\rangle-\xi_s)+z_{ISW}-\langle z_{ISW}\rangle]
\nonumber\\
&& \qquad \times
\left\{ 1  -\frac{1}{2} {1\over\eta_o-\eta_s}\int^{\eta_o}_{\eta_s} (\eta_o - \eta) (\nabla^2 \xi - n^i n^j \xi_{,ij}) (\eta - \eta_s) \mathrm{d}\eta \right\}.
\label{lumdis3b}
\end{eqnarray}
where the different contributions to the redshift, the cosmological, local gravitational, and integrated Sachs--Wolfe effects are:
\begin{equation}
1+z_c = \frac{a_o}{a_s},
\end{equation}
\begin{equation}
1+z_{gr} = \sqrt{\frac{1+2\Psi_o}{1+2\Psi_s}} \approx 1+\Psi_o - \Psi_s,
\end{equation}
\begin{equation}
1+z_{ISW} = 1 - \int^{\eta_o}_{\eta_s} \xi_{,\eta} \mathrm{d}\eta
= 1 - \int^{t_o}_{t_s} \xi_{,t} \mathrm{d}t
\end{equation}
In certain cases, a single equation for $d_L(z)$ can be derived and this equation can be cast as a cosmographic series in $z$. For instance, we showed that for a FLRW universe, we have \eqref{henry2}
\begin{align}
d_L = \frac{a_o}{\mathcal{H}_o}&\left[z - \frac{\mathcal{Q}_o}{2} \, z^2 
+ \frac{3\mathcal{Q}_o^2 + 3\mathcal{Q}_o - \mathcal{J}_o}{6} \, z^3 + O(z^4)\right].
\end{align}
and that for a sinusoidally varying potential the coefficients of this relation are corrected by terms of order $\epsilon$ as in equation \eqref{aragorn}. A few comments regarding the interpretation of our results are in order.

The redshift as written in equation \eqref{redshift2} is a sum of three contributions --- a cosmological redshift, a gravitational redshift, and a redshift due to an ISW effect. However, what we measure only is the total redshift which includes also a Doppler contribution due to the peculiar velocities of the source and the observer. This can be included by hand in the expression \eqref{redshift2} by writing
\begin{equation}
1+z = (1+z_{D})(1+z_{c})(1+z_{gr})(1+z_{ISW}),
\end{equation}
where the Doppler contribution to the redshift is
\begin{equation}
1+z_D = \frac{\gamma_s (1-\vec{v}_s \cdot \hat{n})}{\gamma_o (1-\vec{v}_o\cdot \hat{n})},
\end{equation}
and where $\gamma = (1 - |\vec{v}|^2)^{-\frac{1}{2}}$ and $\vec{v}_s$, $\vec{v}_o$ are the peculiar velocities of the source and the observer. Usually, one assumes that the peculiar velocities of the sources are random and therefore cancel each other out for a large enough sample, while the peculiar velocity of the observer can be canceled from the dipole of the CMB angular distribution~\cite{Planck2}. Within our approach peculiar velocities can be estimated from \eqref{PC1} for FLRW or from \eqref{PC2} for the toy model.

Compared to previous discussions of the luminosity distance in perturbed FLRW Universes such as in~\cite{Durrer} we have made the following improvements. We keep both $\Psi$ and $\Phi$ as general functions of the spacetime coordinates without assuming any relation between them thus keeping our discussion as general as possible within linear perturbation theory. We derive our results using both the Jacobi map and the van Vleck determinant approaches verifying that they give the same results as they should~\cite{Non-perturbative}. While the Jacobi map is extensively used in Cosmology, to the best of our knowledge we are the first to extensively use the van Vleck determinant in the analysis of the luminosity distance. The van Vleck determinant is a mathematical object which appears in many other areas of theoretical physics, and there are multiple techniques to calculate it in certain specific cases of interest~\cite{vanVleck, Visser:1993, Visser:1995}. For current purposes, the van Vleck determinant formalism is mathematically equivalent to the Jacobi determinant formalism but in general the van Vleck determinant has a cleaner physical interpretation in terms of the focussing and defocussing of geodesic flows in a curved spacetime. For that reason it is an important tool in the analysis of the luminosity distance. We focus on the cosmographic approach, which is the best way to test the underlying geometry, by writing the final result for the luminosity distance in the toy model as a generalised cosmographic series. We show how to systematically include peculiar velocities and Doppler redshifts in the cosmographic series both in FLRW and in the toy model. Finally, we rewrite the general formula for the luminosity distance at first order in perturbation theory as much as possible in terms of various contributions to the redshift giving the final formula \eqref{lumdis3b}.

The result for the luminosity distance has limited utility in the vicinity of conjugate points of the congruence of null geodesics emanating from the source. The vector field $\delta x^{\mu}(\lambda)$ is a Jacobi field on the congruence of geodesics and it certainly has a conjugate point at the source: $\delta x_s =0$. If the observer is located at or near another conjugate point, then $\delta x_o \approx 0$, so that $\mathcal{J}(O,S) \approx 0$ and $d_L \approx 0$. For example, if the source and the observer are located on antipodal points in closed FLRW, the luminosity distance between them is zero. Physically this corresponds to the fact that all photons emitted at the source reach the observer, the observer sees the source at all directions in the sky, as if he is located inside the source.

The cosmographic series \eqref{sam1} and \eqref{aragorn}, if formally extended to infinite order,  converge for $|z|<1$ and diverge for $|z|>1$. In order to fit supernovae at higher redshifts, it is useful to perform the cosmographic expansion in terms of the improved parameter $y = \frac{z}{1+z}$~\cite{Cattoen:2007b, Vitagliano}.

Our equations can be applied and the discussion extended in several different directions. The first application is to explore the influence of inhomogeneities from the cosmic structure on the estimation of the cosmographic parameters. The cosmographic parameters are usually estimated by fitting the data from Type Ia supernovae with the theoretical relation \eqref{sam1} which is derived by assuming an ideal FLRW cosmology. Fitting the data with a theoretical relation adapted to an inhomogeneous universe such as \eqref{lumdis3b} might lead to alteration of the estimated values of the Hubble parameter, deceleration parameter and jerk. A second application is to analyse and constrain alternative cosmological models which go beyond the $\Lambda$CDM, for instance, models in which dark energy is dynamical or in which it varies stochastically with cosmic time~\cite{EFTDE, Caldwell, Mukhanov, Everpresent}. However, one has to be careful since our equations are entirely kinematic in nature and insensitive to the precise gravitational dynamics. In order to constrain the deviations from the standard homogenous and isotropic FLRW cosmology, the best approach is to consider supernovae in a tiny shell of fixed size $\Delta z$, at a fixed redshift $z$, and to look at the power spectrum of the luminosity distance. In a completely isotropic cosmology only the monopole would be active and therefore the size of the higher multipole excitations would give a constraint on the possible departures from isotropy. The last application would be to try to put constraints on the values of the peculiar velocities within cosmography. We leave all further investigations along these lines to future work.

%
\acknowledgments
Matteo Viel was partially supported  by the INFN PD51 INDARK grant.\\
Matt Visser was supported by the Marsden Fund, which is administered by the Royal Society of New Zealand.
Matt Visser would like to thank SISSA and INFN (Trieste) for hospitality during the early phase of this work.

\appendix
\section{Demonstration that $\delta \theta_s \bot (k_s, U_s)$ and $\delta x_o \bot (k_o, U_o)$.}
\label{A:1}
Here we demonstrate that the vectors $\delta \theta_s^{\mu}$ and $\delta x_o^{\mu}$ indeed belong to two dimensional subspaces orthogonal to $k^{\mu} = \tilde{\omega} \ell^{\mu}$ and to $U^{\mu}_s, U^{\mu}_o$.

\paragraph{$\delta x_o \bot k_o$:}
Since all photons start at the same point in spacetime, they must have the same phase $P$ defined as
\begin{equation}
\ell_{\mu} = \nabla_{\mu} P.
\end{equation}
Since the phase does not change along a cross section of the congruence, we must have that
\begin{equation}
0=\nabla_{\delta x}P = \delta x^{\mu} \nabla_{\mu} P = \delta x^{\mu}\ell_{\mu},
 \label{once}
\end{equation}
which implies that $\delta x \bot k_o$.

\paragraph{$\delta \theta_s \bot k_s$:}
Define
\begin{equation}
v^{\mu} := \frac{D Y^{\mu}}{d \lambda} = \ell^{\rho} \nabla_{\rho} Y^{\mu},
\end{equation}
so that $\delta \theta^{\mu} = v^{\mu} \delta y$. Then we have that
\begin{equation}
v^{\mu}\ell_{\mu} =\ell_{\mu} \ell^{\rho} \nabla_{\rho} Y^{\mu} = \ell^{\rho} \nabla_{\rho} (l_{\mu} Y^{\mu}) - Y^{\mu} \ell^{\rho} \nabla_{\rho}\ell_{\mu} = 0,
\end{equation}
where the first term vanishes due to \eqref{once} and the second term vanishes due to the geodesic equation. This implies that $\delta \theta_s \bot k_s$.

\paragraph{$\delta \theta_s \bot U_s$:}
This follows from the fact that spacetime at the source $S$ is locally Minkowski and the emission of light is isotropic in all directions.

\paragraph{$\delta x_o \bot U_o$:}
In order for this to hold we must choose a suitable parametrisation of the one-parameter family of null geodesics. Let's say that we start with parameters $(\lambda, y)$ such that $\delta x_o \cdot U_o \neq 0$.
We can obtain new parameters $(\tilde{\lambda}, \tilde{y})$ by performing a general coordinate transformation on the 2-surface spanned by $Y^{\mu}$ and $\ell^{\mu}$
\begin{equation}
\lambda = g_1 (\tilde{\lambda}, \tilde{y});
\end{equation}
\begin{equation}
y = g_2 (\tilde{\lambda}, \tilde{y}).
\end{equation}
However, we want this transformation to preserve the null geodesic curves and to preserve the affinity of the parameter $\lambda$. Thus we are left with
\begin{equation}
\lambda = \tilde{\lambda} + h(\tilde{y});
\end{equation}
\begin{equation}
y = g(\tilde{y}).
\end{equation}
This implies that
\begin{align}
\tilde{Y}^{\mu} &= \frac{\partial f^{\mu}}{\partial \tilde{y}} \nonumber \\
&= \frac{\partial \lambda}{\partial \tilde{y}} \frac{\partial f^{\mu}}{\partial \lambda} + \frac{\partial y}{\partial \tilde{y}} \frac{\partial f^{\mu}}{\partial y} \nonumber \\
&= \frac{\partial h}{\partial \tilde{y}} \ell^{\mu} + \frac{\partial y}{\partial \tilde{y}} Y^{\mu},
\end{align}
which in turn implies
\begin{equation}
\delta \tilde{x}^{\mu} := \tilde{Y}^{\mu} \delta \tilde{y} = \ell^{\mu} \delta h + \delta x^{\mu}.
\end{equation}
Hence
\begin{equation}
\delta \tilde{x}^{\mu}_o U_{O\mu} = (l_{O}^{\mu} U_{O\mu})\delta h + \delta x^{\mu}_o U_{O\mu},
\end{equation}
and this will be zero, provided  
we choose the function $h$ such that
\begin{equation}
\delta h = -\frac{\delta x_o^{\mu} U_{O\mu}}{\ell_o^{\mu} U_{O\mu}}.
\end{equation}

\section{Calculating the Jacobi map and Jacobi determinant}

\label{A:2}

We now show the full calculation of the Jacobi map and Jacobi determinant in the perturbed FLRW spacetime. We first work in the unphysical spacetime \eqref{metr}.  The system of equations
\eqref{de2} reduces to
\begin{equation}
\frac{\mathrm{d}}{\mathrm{d} \hat{\lambda}}(\delta \hat{x}^{(1)\alpha}) = C^{(1)\alpha}_{\nu} (\hat{\lambda}) \; \delta \hat{\bar{x}}^{\nu} + (\delta \hat{\theta}^{\alpha})^{(1)};
\end{equation}
\begin{equation}
\frac{\mathrm{d}}{\mathrm{d} \hat{\lambda}}(\delta \hat{\theta}^{\alpha})^{(1)} = A^{(1)\alpha}_{\nu} (\hat{\lambda}) \; \delta \hat{\bar{x}}^{\nu} + C^{(1)\alpha}_{\nu} (\hat{\lambda}) \; (\overline{\delta \hat{\theta}^{\alpha}}).
\end{equation}
The background equations are the same as those for Minkowski space, 
\eqref{mink2}. Therefore the background unprojected Jacobi map is given by \eqref{unproj}, so
\begin{equation}
\hat{\bar{\mathcal{J}}}^{\alpha}_{\beta} = (\hat{\lambda}_o - \hat{\lambda}_s) \,\delta^{\alpha}_{\beta},
\end{equation}
while the first order correction to the unprojected Jacobi map is given by
\begin{align}
\hat{\mathcal{J}}^{(1)\alpha}_{\beta} &= \int^{\hat{\lambda}_o}_{\hat{\lambda}_s} C^{(1)\alpha}_{\beta} (\hat{\lambda}) (\hat{\lambda} - \hat{\lambda}_s) \mathrm{d} \hat{\lambda} \nonumber \\
&+ \int^{\hat{\lambda}_o}_{\hat{\lambda}_s} \int^{\hat{\lambda}}_{\hat{\lambda}_s} \Big[ A^{(1)\alpha}_{\beta} (\hat{\lambda}^{\prime}) (\hat{\lambda}^{\prime} - \hat{\lambda}_s) + C^{(1)\alpha}_{\beta} (\hat{\lambda}^{\prime}) \Big] \mathrm{d} \hat{\lambda}^{\prime} \mathrm{d} \hat{\lambda},
\end{align}
where
\begin{equation}
C^{(1)\alpha}_{\beta} := - \Gamma^{(1)\alpha}_{\mu \beta} \,\hat{\bar{k}}^{\mu};
\qquad
A^{(1)\alpha}_{\beta} := R^{(1)\alpha}_{\rho \mu \beta} \,\hat{\bar{k}}^{\rho} \hat{\bar{k}}^{\mu}.
\end{equation}
and
\begin{equation}
\hat{\bar{k}}^{\mu} = \tilde{\omega} (1, \vec{n})
\end{equation}
Calculating these for the metric \eqref{metr}, we obtain:
\begin{equation}
C^{(1)0}_{0} = -(\dot{\xi} + \vec{\nabla} \xi . \vec{n});
\qquad
C^{(1)j}_{0} = C^{(1)0}_{j} = -\xi_{,j};
\qquad
C^{(1)j}_{k} = 0;
\end{equation}
and 
\begin{equation}
A^{(1)0}_{0} = \xi_{,ij} n^i n^j;
\qquad
A^{(1)k}_{0} = - A^{(1)0}_k = \xi_{,ki} n^i;
\qquad
A^{(1)k}_{l} = - \xi_{,kl}.
\end{equation}
The photon direction vectors
\eqref{phodir2} can be split into background plus perturbation
\begin{equation}
n^{\mu}_s = \bar{n}^{\mu}_s + n^{(1)\mu}_s;
\qquad
n^{\mu}_o = \bar{n}^{\mu}_o + n^{(1)\mu}_o;
\end{equation}
where
\begin{equation}
\bar{n}^{\mu}_s = \bar{n}^{\mu}_o = (0, \vec{n});
\end{equation}
and
\begin{equation}
n^{(1)\mu}_s = (0, \hat{\vec{\ell}}^{(1)}_s) \approx \vec{0};
\qquad
n^{(1)\mu}_o = (0, \hat{\vec{\ell}}^{(1)}_o).
\end{equation}
Also the projectors
\eqref{proj2} can be split into background plus perturbation:
\begin{equation}
\bar{P}^{\mu}_{s\nu} = \delta^{\mu}_{\nu} + \hat{\bar{U}}^{\mu}_s \hat{\bar{U}}_{s \nu} - \bar{n}^{\mu}_s \bar{n}_{s\nu};
\qquad
\bar{P}^{\mu}_{o\nu} = \delta^{\mu}_{\nu} + \hat{\bar{U}}^{\mu}_o \hat{\bar{U}}_{o \nu} - \bar{n}^{\mu}_o \bar{n}_{o\nu};
\end{equation}
and
\begin{equation}
P^{(1)\mu}_{s\nu} = \hat{\bar{U}}^{\mu}_s \hat{U}^{(1)}_{s\nu} + \hat{U}^{(1)\mu}_s \hat{\bar{U}}_{s\nu} ;
\qquad
P^{(1)\mu}_{o\nu} = \hat{\bar{U}}^{\mu}_o \hat{U}^{(1)}_{o\nu} + \hat{U}^{(1)\mu}_o \hat{\bar{U}}_{o\nu} 
- \bar{n}^{\mu}_o n^{(1)}_{o\nu} - n^{(1)\mu}_o \bar{n}_{o\nu}.
\end{equation}
In terms of the metric \eqref{metr}, the projectors are given by
\begin{equation}
\bar{P}^0_0 = \bar{P}^0_i = \bar{P}^i_0 = 0;
\qquad
\bar{P}^i_j = \delta^i_j - n^i n_j;
\end{equation}
and
\begin{equation}
P^{(1)\mu}_{s\nu} = 0;
\qquad
P^{(1)0}_{o0} = P^{(1)0}_{oi} = P^{(1)i}_{o0} = 0;
\qquad
P^{(1)i}_{oj} = -n^i \hat{\ell}^{(1)}_{oj} - \hat{\ell}^{(1)i}_o n_j.
\end{equation}
The projected Jacobi map is given by
\begin{align}
\hat{J} &= P_o \hat{\mathcal{J}} P_s \\ \nonumber
&= \bar{P}_o \hat{\bar{\mathcal{J}}} \bar{P}_s + P^{(1)}_o \hat{\bar{\mathcal{J}}} \bar{P}_s + \bar{P}_o \hat{\mathcal{J}}^{(1)} \bar{P}_s \\ \nonumber
&= \hat{\bar{J}} + \hat{J}^{(1)}.
\end{align}
The background projected Jacobi map is the same as in Minkowski space
\begin{equation}
\hat{\bar{J}}^0{}_0 = \hat{\bar{J}}^0{}_i = \hat{\bar{J}}^i{}_0 = 0;
\end{equation}
\begin{equation}
\hat{\bar{J}}^i{}_j = (\hat{\lambda}_o - \hat{\lambda}_s) (\delta^i{}_j - n^i n_j).
\end{equation}
After a long but straightforward calculation, the full projected Jacobi map can be shown to be equal to
\begin{equation}
\hat{J}^0{}_0 = \hat{J}^0{}_i = \hat{J}^i{}_0 = 0;
\end{equation}

\begin{align}
\hat{J}^i{}_j &= (\hat{\lambda}_o - \hat{\lambda}_s) \Big[ \delta^i{}_j - n^i n_j \Big] \nonumber \\
&- \int^{\hat{\lambda}_o}_{\hat{\lambda}_s} \int^{\hat{\lambda}}_{\hat{\lambda}_s} \Big[\xi_{,ij} - \xi_{,il} n^{,l} n_{,j} - n^i n_k \xi^{,k}_{,j} + n^i n_k \xi_{,kl} n^l n_j \Big] (\hat{\lambda}^{\prime} - \hat{\lambda}_s) \mathrm{d}\hat{\lambda}^{\prime} \mathrm{d} \hat{\lambda} \nonumber \\
&+ (\hat{\lambda}_o - \hat{\lambda}_s) \Big[ -n^i \hat{\ell}^{(1)}_{oj} + n^i n_j (\hat{\vec{\ell}}^{(1)}_o\cdot \vec{n}) \Big]. \label{jones}
\end{align}
Solving the characteristic equation, we find the two non-vanishing eigenvalues of this $3\times3$ spatial matrix. Their product gives the determinant of the Jacobi map (to first order)
\begin{align}
(\det \hat{J})^{\frac{1}{2}} &= (\hat{\lambda}_o - \hat{\lambda}_s) 
 -\frac{1}{2} \int^{\hat{\lambda}_o}_{\hat{\lambda}_s} \int^{\hat{\lambda}}_{\hat{\lambda}_s} (\hat{\lambda}^{\prime} - \hat{\lambda}_s) \xi_{,ij} (\delta_{ij} - n_i n_j) \mathrm{d}\hat{\lambda}^{\prime} \mathrm{d}\hat{\lambda}. \label{smith}
\end{align}
We can rewrite the double integral as a single integral by using the identity
\begin{equation}
\int^{\eta_o}_{\eta_s} \int^{\eta}_{\eta_s} g(\eta^{\prime}) \mathrm{d} \eta^{\prime} \mathrm{d} \eta = \int^{\eta_o}_{\eta_s} (\eta_o - \eta) g(\eta) \mathrm{d} \eta.
\end{equation}
We then obtain
\begin{align}
(\det \hat{J})^{\frac{1}{2}} &= (\hat{\lambda}_o - \hat{\lambda}_s)   \left\{ 1 
-\frac{1}{2} {1\over \hat{\lambda}_o - \hat{\lambda}_s} \int^{\hat{\lambda}_o}_{\hat{\lambda}_s} (\hat{\lambda}_o - \hat{\lambda}) (\nabla^2 \xi - n^i n^j \xi_{,ij}) (\hat{\lambda} - \hat{\lambda}_s) \mathrm{d}\hat{\lambda}\right\}.
 \label{jacdet2}
\end{align}
Since $(\hat{\lambda}_o - \hat{\lambda}_s)$ is positive and $\xi$ is by assumption extremely small we must have that $(\det \hat{J})^{\frac{1}{2}} = |\det \hat{J}|^{\frac{1}{2}}$.
If one desires, one can now easily obtain the Jacobi map and Jacobi determinant in the full perturbed FLRW spacetime by using the relations \eqref{conf1} and \eqref{conf2}:
\begin{equation}
J = f_o\; \hat{J}; \qquad\hbox{and} \qquad \det J = f_o^2\; \det \hat{J}.
\end{equation}


\bibliography{Bibtex}	

\end{document}